\documentclass[twocolumn, superscriptaddress]{revtex4-1}
\usepackage{latexsym}
\usepackage{amsmath}
\usepackage{amsfonts}
\usepackage{amssymb}
\usepackage{amsbsy}
\usepackage{bm}
\usepackage{color}
\usepackage{graphicx}
\usepackage{longtable}
\usepackage{setspace}
\usepackage{color}

\usepackage{soul}
\usepackage{appendix}
\usepackage{gensymb}
\usepackage{ulem}
\usepackage{tikz}
\usepackage{pict2e}
\usepackage[hidelinks]{hyperref}

\providecommand{\Drw}{$D_r^w$}

\DeclareRobustCommand{\crossedsquare}{%
    \setlength{\unitlength}{0.5ex} 
    \begin{picture}(3,3)
        \put(0,0){\framebox(3,3){}} 
        \put(1.5,0){\line(0,1){3}}  
        \put(0,1.5){\line(1,0){3}}  
    \end{picture}%
}

\DeclareRobustCommand{\halffilledsquare}{%
    \tikz[baseline=-0.1ex, scale=0.2]{
        \draw (0,0) rectangle (1,1); 
        \fill[black] (0,0) rectangle (1,0.5); 
    }%
}

\begin{document}

\begin{titlepage}
\title{
Active wetting transitions induced by rotational noise at solid interfaces
}

 \author{Suchismita Das}
 \thanks{Current address: Laboratoire de Physique de l'École normale supérieure, Université PSL, CNRS, Sorbonne Université and Université Paris Cité, 75005 Paris, France}
 \affiliation{Department of Physics, Indian Institute of Technology Bombay, Mumbai - 400076, India.}
 \affiliation{Max Planck Institute for the Physics of Complex Systems, N\"{o}thnitzerst. 38, 01187 Dresden, Germany}

 \author{Raghunath Chelakkot}
 \email{raghu@phy.iitb.ac.in}
 \affiliation{Department of Physics, Indian Institute of Technology Bombay, Mumbai - 400076, India.}
\date{\today}


\begin{abstract}

We investigate the wetting transitions displayed by the collection of active Brownian particles (ABPs) confined within rigid, impenetrable, flat walls. In our computational study using Brownian dynamics simulations, the wall-particle interactions are implemented with a short-range repulsive potential. Our analyses reveal that an enhanced rotational diffusion at the walls can be used as a control parameter for wetting transitions in the dense aggregates of active particles at the wall. 
Increasing the wall rotational diffusion destabilizes a uniform, complete wetting state, and the aggregate shows morphological transitions. We observe a sequence of morphological transitions with an increase in wall rotational diffusion: symmetric complete wetting (SCW), asymmetric complete wetting (ACW), partial wetting (PW) with droplet formation, and drying. We compute the contact angle in the PW state as a function of activity and rotational noise. Our analysis indicates that these transitions are linked to enhanced kinetic energy fluctuations of particles and bubble formations in the dense state.  
We further characterize the nature of these transitions by systematically analyzing an order parameter. Our work shows that modifying local reorientation rates alone is sufficient to induce wetting transitions in active systems.
\end{abstract}

\maketitle
\end{titlepage}

\section{Introduction}

The spreading of a liquid on a solid substrate has garnered significant research attention across diverse fields over the past several decades~\cite{Bonn_2009, deGennes_RMP_1985, gennes2004capillarity}. When a liquid drop is placed on a solid substrate, the equilibrium state is described by a coexistence of three phases and interfaces. The morphology of the liquid phase is determined by the relative strengths of the surface tensions. A wetting transition occurs when the liquid drop spreads on the substrate, forming a macroscopic layer of liquid. 
These transitions are studied extensively using theory~\cite{cahn1977critical, Pandit_PRB_1982, nakanishi1982multicriticality, dietrich1991analytic, lipowsky1986wetting}, experiments~\cite{moldover1980interface,pohl1983wetting,bonn1992hysteresis, Douezan_SM_2012,Edwards_Sci_Adv,murata2010surface,hejazi2012wetting,bussonniere2017universal,yu2023wetting,lambley2023freezing} and simulations~\cite{nijmeijer1990wetting, vanRemoortere_JCP_1999}. The morphological state of the liquid phase is typically characterized by the contact angle formed by the liquid phase on the solid substrate. The contact angle is zero for a complete wetting state, whereas for a partial wetting state (liquid drop), it has a value between zero and $\pi/2$. When the contact angle is larger than $\pi/2$, the state is termed non-wetting ~\cite{Bonn_2009, gennes2004capillarity}. In a partial wetting state, the macroscopic contact angle and three different surface tensions are related by the famous Young-Dupr\'e equation. The complete drying state is denoted by the absence of a solid-liquid interface. Both theoretical and experimental studies have revealed that wetting behavior is affected by the nature of solid-fluid interaction and temperature. 

While the properties of passive fluids on solid substrates have been extensively studied, there is a growing interest in exploring the aggregation and spreading of several active and biological systems on solid substrates and membranes~\cite{Douezan_PNAS_2011, Perez-Gonzalez_NatPhys_2019,pallares2023stiffness,Kusumaatmaja_JCB2021,Agudo-Canalejo2021,Alert_Langmuir_2019,Zhao_2021,melaugh2016shaping, ponisch2019bacterial,fins2024steer,Sandor_JCP_2017,reichhardt2014absorbing}. 
In addition, the spread of bacterial species on substrates has also been studied in this context~\cite{melaugh2016shaping, ponisch2019bacterial}. Unlike their passive counterparts, such living systems continuously convert energy into mechanical motion and hence far from equilibrium. The collective behavior of such self-motile systems has been theoretically studied using minimal particle-based models. It is well known that such systems display bulk phase separation at sufficiently large density and motility, called Motility-induced phase separation (MIPS)~\cite{Redner2013, Fily2012, Cates2015}. They also display wall accumulation under confinement, even in the absence of explicit adhesive interactions~\cite{Bechinger2016, lee2017interface, Elgeti2013, caprini2018active}. 
Several numerical studies have investigated the nature of wall aggregation of self-propelling particles~\cite{sepulveda2017wetting, SepulvedaSoto_PRE2018_activewetting, neta2021wetting, Turci_PRL_2021, Das2020_SM, Das2020_PRE, Turci_SM_2024}. It has been shown that the particle aggregates display a continuous transition from a complete wetting to a partial wetting state on rigid, impenetrable walls as the particle reorientation rate is increased~\cite{sepulveda2017wetting, SepulvedaSoto_PRE2018_activewetting}. Such wetting transitions of active aggregates are also observed when the substrate properties are altered. The effect of wall permeablilty~\cite{Turci_PRL_2021, Turci_SM_2024} and porosity~\cite{Das2020_SM, Das2020_PRE}, and particle inertia~\cite{caprini2024dynamical} are explored in some detail. {Further, a mechanical definition of liquid-solid surface tension has been derived recently~\cite{zakine2020surface, zhao2024active}}. An intriguing morphological transition is observed in the aggregates of particles on the porous walls. With a slight increase in the particle crossing rate, the aggregates display a morphological transition from a uniformly spread state to a liquid droplet with a non-zero contact angle~\cite{Das2020_SM,Das2020_PRE}, displaying a striking similarity to equilibrium wetting transitions. Apart from systems of self-propelling particles, studies have also been conducted on wetting transitions in other non-equilibrium models~\cite{Hinrichsen_PRL_1997, Hinrichsen_PRE_2003}.

All the above numerical studies on wall aggregation of active systems are based on the assumption that the intrinsic properties of the active particle, such as motility and reorientation rate, are independent of external conditions. However, in several bacterial systems, it is known that the individual behavior of organisms can be altered in response to external stimuli~\cite{whiteley2017progress}.
Further, it is also known that bacterial motility behavior is modified in the vicinity of rigid walls~\cite{Junot_PRL_2022, tokarova2021patterns}.  Bacterial suspensions including {\textit{Pseudomonas aeruginosa}} and \textit{Esherichia coli} \cite{Sartori_PRE_2018, Bianchi_PRX_2017} have also exhibited altered behavior when confined within rigid walls. Under confinement, {\textit{Pseudomonas aeruginosa}} shows a run and reverse motion \cite{Sartori_PRE_2018}, while {\textit{Esherichia coli}} cells experience reorientation only when in contact with a flat wall \cite{Bianchi_PRX_2017}. Also, in {\it Escherichia coli}, their ability to actively modify reorientation rates of their propulsion direction in response to chemical gradients is well characterized~\cite{tu2013quantitative, colin2017emergent, colin2021multiple, waite2018behavioral}.
This naturally leads to a question of how the wall accumulation and the wetting behavior of active agents are influenced by the modification of their reorientation in the vicinity of rigid boundaries.  A generic model of active particles in which individual agents modify their reorientation rate while in contact with a rigid wall will potentially help to understand the active wetting properties observed in various experimental systems.

In this work, we report wetting transitions for active particles in the presence of rigid, repulsive walls. In previous work on active particles in the presence of porous walls \cite{Das2020_SM, Das2020_PRE}, the morphological transition has been studied as a function of the change in wall porosity and particle motility. It has been shown that a slight increase in the pore size enhances the cross-wall particle flux, leading to a significant change in the pressure distribution of the dense particle clusters, which eventually amounts to a change in the morphology of particle aggregates.  
A detailed analysis gives us a cue that altering wall-particle interactions by enhancing the noise at the wall might lead to wetting transitions even in the presence of rigid, 
impenetrable walls without explicitly changing the interaction potential. 

 In numerical models, enhanced reorientation rate at the walls can be introduced in several ways. For instance, the noise can be enhanced by increasing the tumbling rate in the case of run-and-tumble particles. However, here, we use a system of active Brownian particles (ABP), where the reorientation is introduced in the form of enhanced rotational diffusion near the walls. The bulk phase properties of ABP systems have been extensively studied as a function of motility, density, interaction strength, particle composition, etc.~\cite{Redner2013,sanoria2021influence, sanoria2024percolation, sanoria2022percolation, hopkins2023motility}. These 
 extensive studies on ABP models can act as reliable benchmarks for the results obtained by systematically changing the noise at the wall. 
\section{Model and simulation details}

We consider a two-dimensional system of $N$ active Brownian particles (ABPs) with diameter $\sigma$, distributed in a rectangular box with dimensions $L_x$ and $L_y$. The system is confined in the $y$ direction  by means of  rigid, impenetrable walls at $y = 0$ and $y = L_y$.
We also introduce periodic boundary conditions along the $x$ direction.  Both the inter-particle and wall-particle interactions are repulsive, derived via the Weeks-Chandler-Andersen (WCA) \cite{WCA1971} potential $U$. The inter-particle interaction force $\textbf{F}_i$ as well as the wall-particle interaction force $\textbf{F}_i^{\text{w}}$ is given by $-\bm{\nabla}_i U$. 
\par
The dynamics of the position of the active particle ${\bf{r}}_i$ follows the overdamped equation
\begin{equation}
{\dot{\bf{r}}}_i  = D\beta\left(\textbf{F}_i + \textbf{F}^{\text{w}}_i \right) + v_0{\hat{\bf{e}}}_{i}+ \sqrt{2D}{\bm{\eta}}_i, 
\end{equation}
where $D$ is the translational diffusion coefficient, $\beta = {1 / k_BT}$, and $\eta$ is Gaussian white noise such that $\langle \eta(t) \rangle$ $= 0$ and $\langle \eta_{i\alpha}(t)\eta_{j\beta}(t^\prime)=\delta_{ij}\delta_{\alpha\beta}\delta(t-t^\prime)$. The active particle self-propels with speed $v_0$ along the polarity vector $\hat{\bf e}_i = ( \cos\phi_i, \sin\phi_i)$, where $\phi_i$ is the 
orientation angle for a particle $i$. In usual ABP models, the evolution of the particle orientation $\phi_i$ is governed by a constant rotational diffusion coefficient $D_r$. However, in distinction to this usual evolution rule, we consider a rotational diffusion of the particles to be a function of distance from the walls. In this study, the polarity directions of the ABPs evolve as $\dot{\phi}_{i} = \sqrt{2(D_{r}(y)}\eta_{i}^{R}$, 
where, $D_r(y) = D_{r}+D_r^w[\Theta(r_{c}-|y-y_{w1}|) + \Theta(r_{c}-|y-y_{w2}|)]$, and $\eta_R$ is the Gaussian white noise with zero mean and unit variance. Here, $\Theta$ is the Heaviside step function with the $y$ separation between the particle positions and the wall positions ($y_{w1} =0$ and $y_{w2}=L_y$) as argument. Thus, the particles have an added rotational diffusion ($D_r^w$) if their distance from either of the walls is less than $r_c$. Here, we take $r_c = 2^{1/6}\sigma$, the WCA interaction cut-off. Away from the wall, the particles reorient according to the bulk rotational diffusion coefficient, $D_r = 3D/\sigma^2$. 
The non-dimensional units of time and energy here are measured as $t_0 = {\sigma^2}/{D}$ and $k_BT$, respectively. The activity is quantified by a nondimensional P{\'e}clet number $Pe = {v_0 \sigma}/{D}$. In our simulations, we choose the $L_y$ to be greater than the persistence length $L_p = v_0/D_r$ of the active particles.
\par
We have conducted the simulations with system sizes varying from $N = 4455$ to $N = 38400$, 
while keeping the area fraction $\varphi = N\pi\sigma^2/4L_xL_y$ fixed at $\varphi \simeq 0.3$. {At this $\phi$, the system is already in the liquid-gas coexistence region, as shown in the previous studies on ABP models with 2D periodic boundary conditions~\cite{siebert2018critical, Bialke2015, levis2017active}. However, the probability of a spontaneous phase separation in the bulk is still very low at this density even for large Pe~\cite{Redner2013,levis2017active}. Under confinement, the phase separation is expected to occur only at the walls.} Most of our simulations are conducted for $N=38400$ in a rectangular box with $L_x/\sigma = 617$ and $L_y/\sigma = 162$. The simulations were run up to time $T/t_0 =3100$, 
for a range of $Pe = 0-300$, maximum up to $10$ independent runs. 
For simulations, we use a time step $10^{-5} t_0$ for $Pe < 180$ and $5 \times 10^{-6} t_0$ for $Pe \geq 180$. 
We vary the rotational diffusion near walls $D_r^w$ in the range of $0 - 4335$. The upper limit of $D_r^w$ is limited such that the time-step of integration is at least an order lesser than the time to rotate for escaping a flat surface $\pi/2D_r^w$. {A summary of the parameters used can be found in 
Table I.}
\newline
{We have not observed a bulk phase separation for any of the parameter values. However, the dense aggregation on walls shows variety of morphological states as we vary $D_r^w$ for different Pe.} 
\section{Results}
\begin{figure}
    \includegraphics[width=0.98\columnwidth]{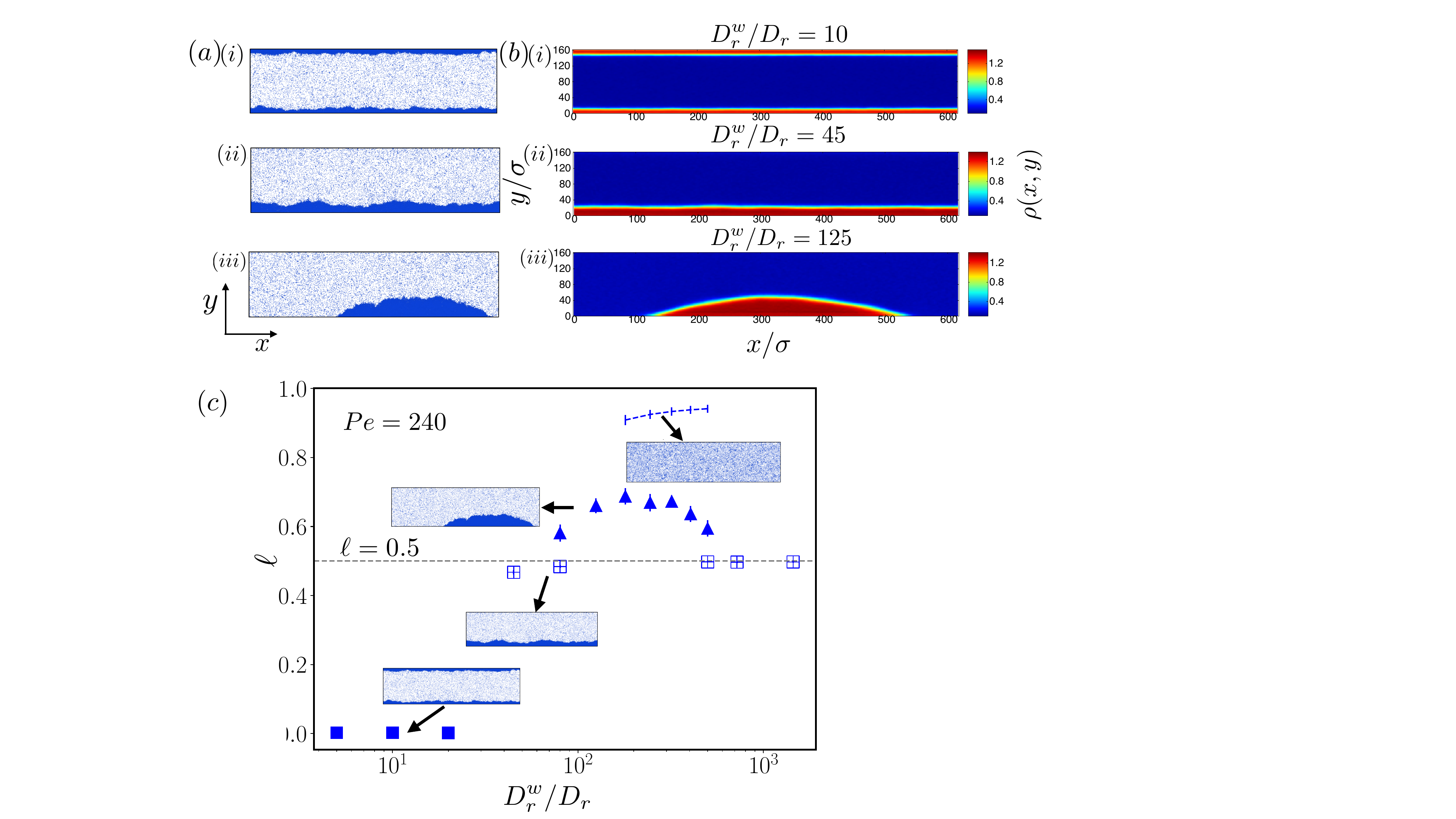}
    \caption{
        {(a)} Examples of cluster morphology with rigid, impenetrable walls for $N =38400$ and $Pe = 240$: 
        {(i)} symmetric complete wetting ($D_r^w/D_r = 10$), 
        {(ii)} asymmetric complete wetting $\crossedsquare$ ($D_r^w/D_r = 45$), 
        {(iii)} partial wetting ($D_r^w/D_r = 125$) (b) 
         Two-dimensional steady-state density distribution, with a color bar denoting the local density $\rho(x,y)$ of the states in {(a)}. 
        {(c)} Vacancy fraction $\ell$ as a function of $D_r^w$ for five independent runs. The error bars denote the standard deviation over time. The symbols denote symmetric complete wetting (SCW)($\blacksquare$), asymmetric complete wetting (ACW)($\crossedsquare$), partial wetting (PW)({\scalebox{1.2}{$\blacktriangle$}}), and complete drying state (CD) (--)
        Here, $N=38400$, $Pe = 240$. (See Supplementary Movies 1-3.)
    }
    \label{fig:1}
\end{figure}

The simulations reveal different steady-state morphologies of dense aggregates on the wall, which are determined by Pe and \Drw. 
When $D_r^w = 0$, the rotational diffusion of ABPs is the same in the bulk and at the wall $(D_r(y) = D_r)$. In this limit, we observe statistically uniform dense layers of particles on both walls when $Pe$ is sufficiently large (Fig.~\ref{fig:1}(a)-i). 
For a given \Drw, the average thickness of such aggregates in such a system is also determined by $\varphi$ and Pe. For $\varphi=0.3$, the thickness of these aggregates is at least an order of magnitude larger than the particle diameter. Also, the average thickness is the same for both walls, which is consistent with previous observations in confined ABP systems~\cite{Das2020_SM, SepulvedaSoto_PRE2018_activewetting, yang2014aggregation}.  
 Thus, the aggregates form a macroscopic film of a high-density phase, resembling complete wetting observed in equilibrium fluids (see supplementary MOVIE1). Here, the dense layer is equivalent to a liquid state, and the dilute phase away from the wall is similar to a vapor state. Therefore, this state is called {\it symmetric complete wetting} (SCW) state. Note that the system does not form a wall-vapor interface in this state.

 When \Drw is increased, keeping the Pe same, particles no longer aggregate symmetrically on both walls; instead, we observe a uniform aggregation on one of the walls and no aggregation on the opposite wall (see Fig.~\ref{fig:1}(a)(ii)). In this state, one of the liquid-vapor interfaces formed in SCW is replaced by a wall-vapor interface. Although a simultaneous occurrence of all three interfaces (liquid-vapor, wall-liquid, and wall-vapor) is a feature of a partial wetting state, the wet wall is covered by a uniform, dense layer of particles, unlike what was typically observed in equilibrium partial wetting (see supplementary MOVIE2). We also observe that the thickness of such a dense layer is approximately twice that of the aggregate thickness formed in SCW and the fraction of particles in the dense state is approximately the same. We term this state, \textit{{asymmetric complete wetting}} (ACW). Such asymmetric states have previously been observed in simulations of Lennard-Jones fluids~\cite{nijmeijer1990wetting, sikkenk1988simulation}. {We have measured the local density of the dense and dilute regions after the phase separation and compared it with the co-existence densities of the bulk systems (see 
 Fig.~S8).
 While the vapour phase density is close to the bulk co-existence density, the liquid phase density is much higher compared to the bulk systems, as the presence of a rigid wall leads to an enhanced compression of the dense state.}
 
 Further increasing $D_r^w$, we observe a qualitatively different aggregate morphology of ABPs. Instead of a uniformly thick dense layer along the wall, the particles form a macroscopically curved liquid-vapor interface on one of the walls, resembling a droplet, while the opposite wall is completely dry (Fig.~\ref{fig:1}(a)(iii))(also, see supplementary MOVIE3). The form of such aggregate is similar to the clusters formed by highly motile active particles on porous walls \cite{Das2020_SM}. Since such dense states resemble liquid droplets formed on solid surfaces in equilibrium systems, we call this state to be \textit{{partial wetting}} (PW). For even higher $D_r^w$, we also observe {\textit{complete drying}} (CD), where dense aggregates are not observed in any of the walls. 
However, our simulations also show that the ACW state reappears for even larger values of $D_r^w$, indicating a re-entrant wetting. 

{For both ACW and PW, the particle condensation is initiated at different locations on the wall for different independent realizations. Once such a condensed layer is formed on a particular wall, it does not switch between the walls during the entire simulation time. However, when a droplet is formed in the PW case, it can gradually change its location along the wall as evident in MOVIE3. However, we observe a completely different behaviour in the absence of interparticle interaction 
(Fig.~S7).
Since interactions are necessary for a phase separation, we do not observe a thick layer of particles on the wall. Therefore, we also do not observe morphological transitions in such cases.}

\subsection{Density profile and vacancy fraction}
Although the instantaneous steady-state configurations clearly indicate a morphological transition of the dense states, we also calculate the time-averaged density distribution $\rho(x,y)$ of ABPs for the characterization of these states (see {Supplementary Material} for details).  As shown in Fig.~\ref{fig:1}(b)(i)-(iii), $\rho$ provides a clear distinction between liquid and vapor states and relatively smooth liquid-vapor interfaces for all the morphological states. While SCW and ACW states provide a flat liquid-vapor interface, the PW state shows a curved interface.  

We further calculate the vacancy fraction $\ell$, which quantifies the fraction of the length of the wall where the dense layer of particles is absent. In other words, $\ell$ quantifies the fractional length of the wall-vapor interfaces  (see {Supplementary Material} for details). The value of $\ell$ is an indicator of the type of morphological state of the dense aggregates. If $\ell \simeq 0$, the system is in SCW state. In the case of ACW, where one of the walls is dry, $\ell \simeq 0.5$. For partial wetting (droplets) $0.5<\ell<1$ and for completely dry state $\ell \simeq 1$. 

In Fig.~\ref{fig:1}(c), we plot $\ell$ as a function of {$D_r^w/D_r$} for a fixed Pe. For low {$D_r^w/D_r$ ($D_r^w/D_r \leq 20$)}, $\ell$ is very low ($\ell < 0.01)$) since both the walls are completely wet, corresponding to the SCW state. At larger \Drw, there is a significant increase in $\ell$, corresponding to the formation of the ACW state. This state is observed in the range {$D_r^w/D_r$ ($45 - 80$)}, where $\ell \sim 0.5$. When the dense state forms a droplet for the range {$80 < D_r^w/D_r < 180$}, $\ell$ gradually increases. In this regime, there are also indications of bi-stability where we observe partial wetting and complete dry states for the same parameter values for different simulations. In Fig.~\ref{fig:1}(c), this is reflected in the range {$180< D_r^w/D_r <500$} where two distinct values of $\ell$ exist for same {$D_r^w/D_r$}. 
At very high values of {$D_r^w/D_r$}, (here {$D_r^w/D_r = 500$} and higher), we find re-emergence of the ACW state. Although Fig.~\ref{fig:1} is for only Pe=240, the trend  SCW$\rightarrow$ACW$\rightarrow$PW$\rightarrow$CD, followed by a re-entrant ACW, is observed for a range of other Pe values.

\subsection{State diagram}
\begin{figure}
    \includegraphics[width=0.98\columnwidth]{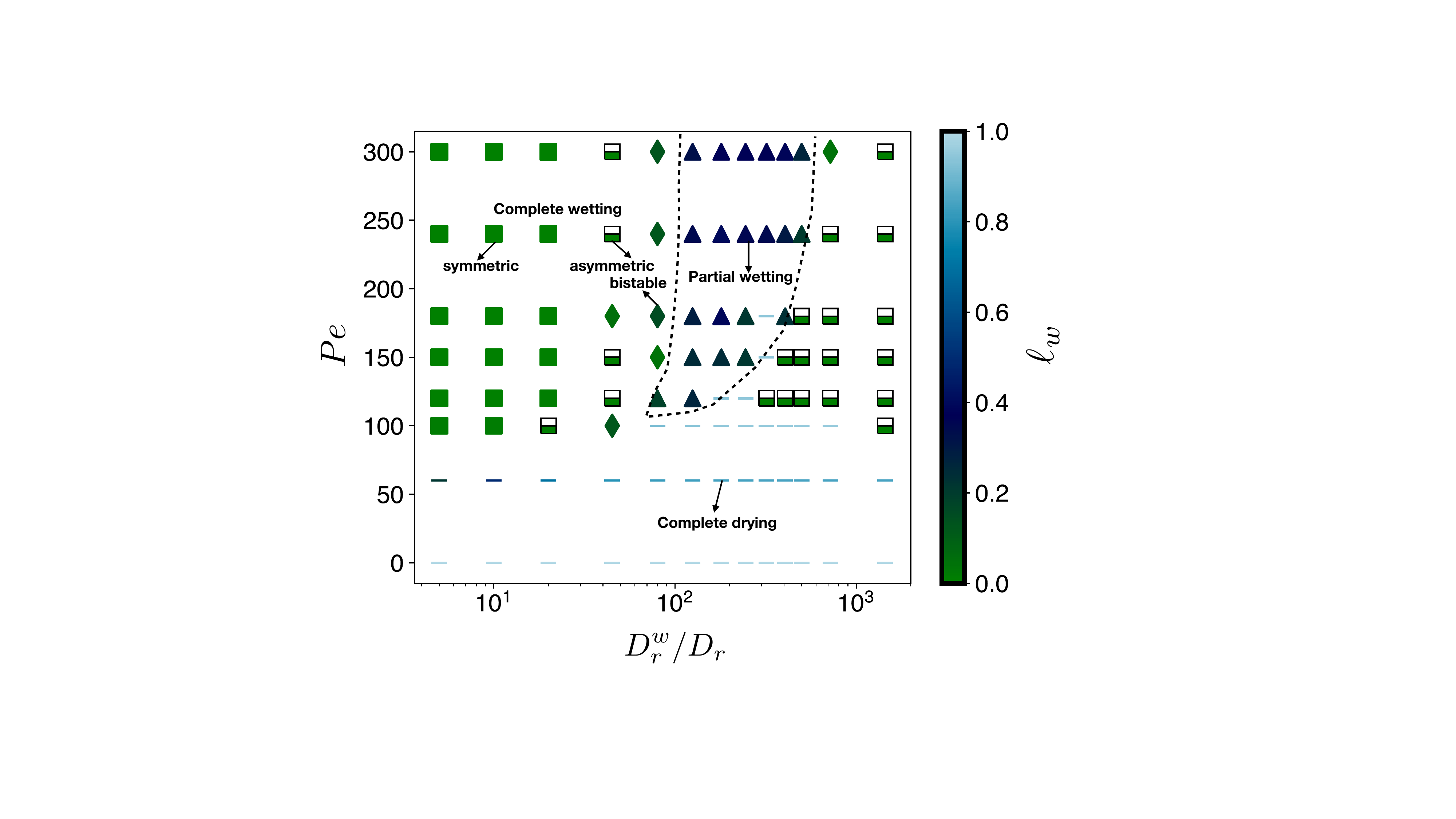}
    \caption{
        State diagram showing steady-state morphology of dense clusters with the variation of activity $Pe$ and {scaled} enhanced rotational diffusion near the wall $D_r^w/D_r$ for the number of active particles, $N = 38400$.
        The color coding in the panel indicates the wet-wall vacancy fraction, $\ell_w$ considered for the wet wall cases, except for states showing complete drying {for all realizations. The system exhibits symmetric complete wetting $\blacksquare$ (Fig.\ref{fig:1}(a)(i)), asymmetric complete wetting $\halffilledsquare$ (Fig.\ref{fig:1}(a)(ii)), partial wetting (Fig.\ref{fig:1}(a)(iii)) {\scalebox{1.2}{$\blacktriangle$}}, with $\blacklozenge$ depicting bistable states between asymmetric and partial wetting, and complete drying states $--$.} 
    }
    \label{fig:2}
\end{figure}

The morphological states of the dense aggregates shown in Fig.~\ref{fig:1} are for a fixed activity ($Pe = 240$). However, the qualitative behavior of the system is also influenced by the particle activity. In Fig.~\ref{fig:2}, we summarize the states as a function of $Pe$ and {$D_r^w/D_r$}. A novel feature observed in the simulation is the formation of the droplet in the PW state, making it crucial to identify the parameter range where partial wetting occurs.  Therefore, we introduce a slightly different definition for vacancy fraction, called the wet-wall vacancy fraction $\ell_w$, which is calculated only for walls that are partially or completely wet. With this definition, both SCW and ACW are treated as morphologically equivalent states with the same $\ell_w$. Thus, ($\ell_w \simeq 0$) for both SCW and ACW and for PW state, $\ell_w >0$. However, we keep track of ACW and SCW by estimating $\ell$ and summarize this information in Fig.~\ref{fig:2}.

 For $Pe \lesssim 60$, we do not observe any aggregation, and the system is in a completely dry state for all {$D_r^w/D_r$}. For higher activity ($Pe \gtrsim 100$), we observe SCW states for low {$D_r^w/D_r$} ({$D_r^w/D_r$} $\lesssim 33.3$) and other aggregated states at higher {$D_r^w/D_r$}. The droplet formations in PW states are observed for a range of {$D_r^w/D_r$} for sufficiently large Pe, which is distinguished by a higher $\ell_w$. This range becomes narrower as $Pe$ is decreased until it disappears for  $Pe \lesssim 100$. This region is also marked by the existence of bi-stable states, where PW states coexist with ACW states. In the state diagram Fig.~\ref{fig:2}, we observe the appearance of re-entrant wetting behavior for {$D_r^w/D_r \geq 500$}. 

 Our analysis indicates that the re-entrance of ACW at high $D_r^w$ is due to the fast reorientation of particles near the walls. For a particle that is oriented away from the wall to escape to the bulk, the minimum escape time ($\sim \sigma/v_0$) should be smaller compared to the typical reorientation time $(D_r^w)^{-1}$, that is, the particle orientation should not change significantly at least during the minimum time which a particle will take to move away from the wall. But, for very high \Drw, the particle reorientation time becomes smaller compared to the escape time; thus, the particle stays near the wall. This leads to a reappearance of a continuous dense layer.    

\subsection{Contact angle of droplets}
\begin{figure}[h]
	\includegraphics[width=0.98\columnwidth]{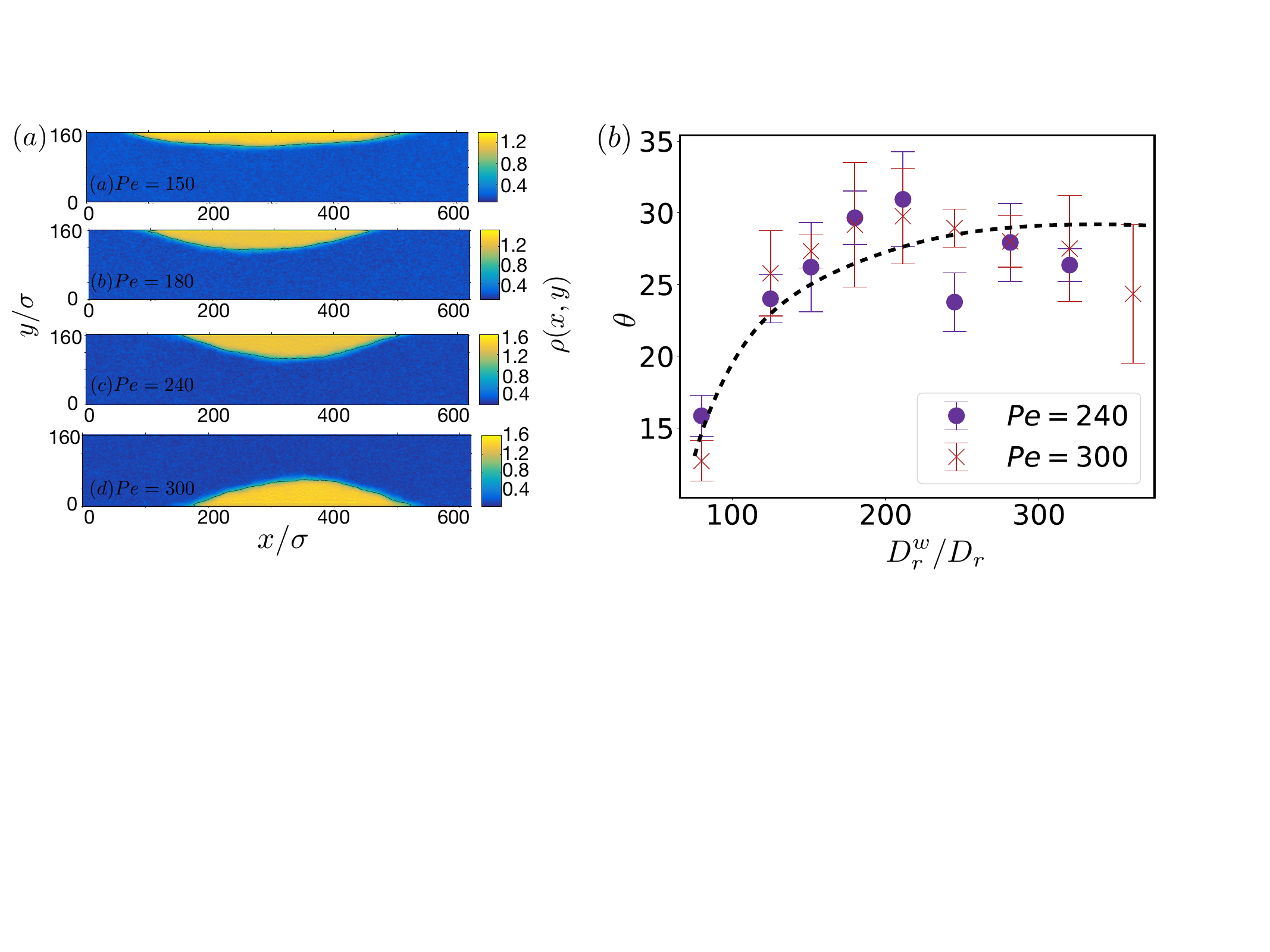}
	\caption{ 
    (a) Time-averaged density profile for droplets keeping $D_r^w/D_r = 180$ for increasing $Pe$, showing gradual increase in curvature of droplets and the contact angle. (b) Contact angle $\theta$ (in $\degree$) for droplets for increasing $D_r^w/D_r$ for $Pe = 240$ and $Pe =300$. The dotted lines are guide to the eye.}
	\label{fig:3}
\end{figure}

Once we identify the parameter range where the droplet states are observed, we analyze how the droplet morphology is influenced by both $Pe$ and \Drw. For this, we use the time-averaged density profile, as it clearly identifies the liquid-vapor interface and its shape. First, we note that for a fixed \Drw, increasing activity leads to a change in droplet shape, as illustrated in Fig.~\ref{fig:3}(a). For small $Pe$, the droplet is relatively flat upon increasing the $Pe$, and the droplet curvature increases until it saturates at large $Pe$ (Fig.\ref{fig:3}(a)). We also calculate the contact angle $\theta$ from the local tangent of the liquid-vapor interface at the liquid-wall contact point for different parameters. The contact angle is low for a relatively flat aggregate, and it increases as the interfacial curvature increases 
{(see Fig.~S4)}
In Fig.~\ref{fig:3}(b), we plot $\theta$ as a function of {$D_r^w/D_r$} for a fixed $Pe$.  Similar to the trend observed with the change in $Pe$, we find that $\theta$ is sensitive to an increase in {$D_r^w/D_r$} for relatively small values of {$D_r^w/D_r$}, and it saturates around $\pi/6$ at large {$D_r^w/D_r$}. We note that we never observe $\theta$ approaching $\pi/2$ for any of the simulation parameters. 

\subsection{Analysis of asymmetric wetting transition}
Our simulations indicate three morphological transitions with a gradual increase in \Drw. These transitions involve partial or complete replacement of the liquid-vapor interface with the wall-vapor interface. One intriguing question is how an increase in \Drw, affecting only those particles in a thin layer in contact with a wall, causes a destabilization of the entire dense layer.  To understand this phenomenon, it is important to analyze the change in properties of the entire dense layer, with an increase in \Drw. We focus on the SCW to ACW transition, as the morphological equivalence of these dense states also allows a close comparison between the structural properties of these states. 

\begin{figure}
 \begin{center}
    	\includegraphics[width=1.0\columnwidth]{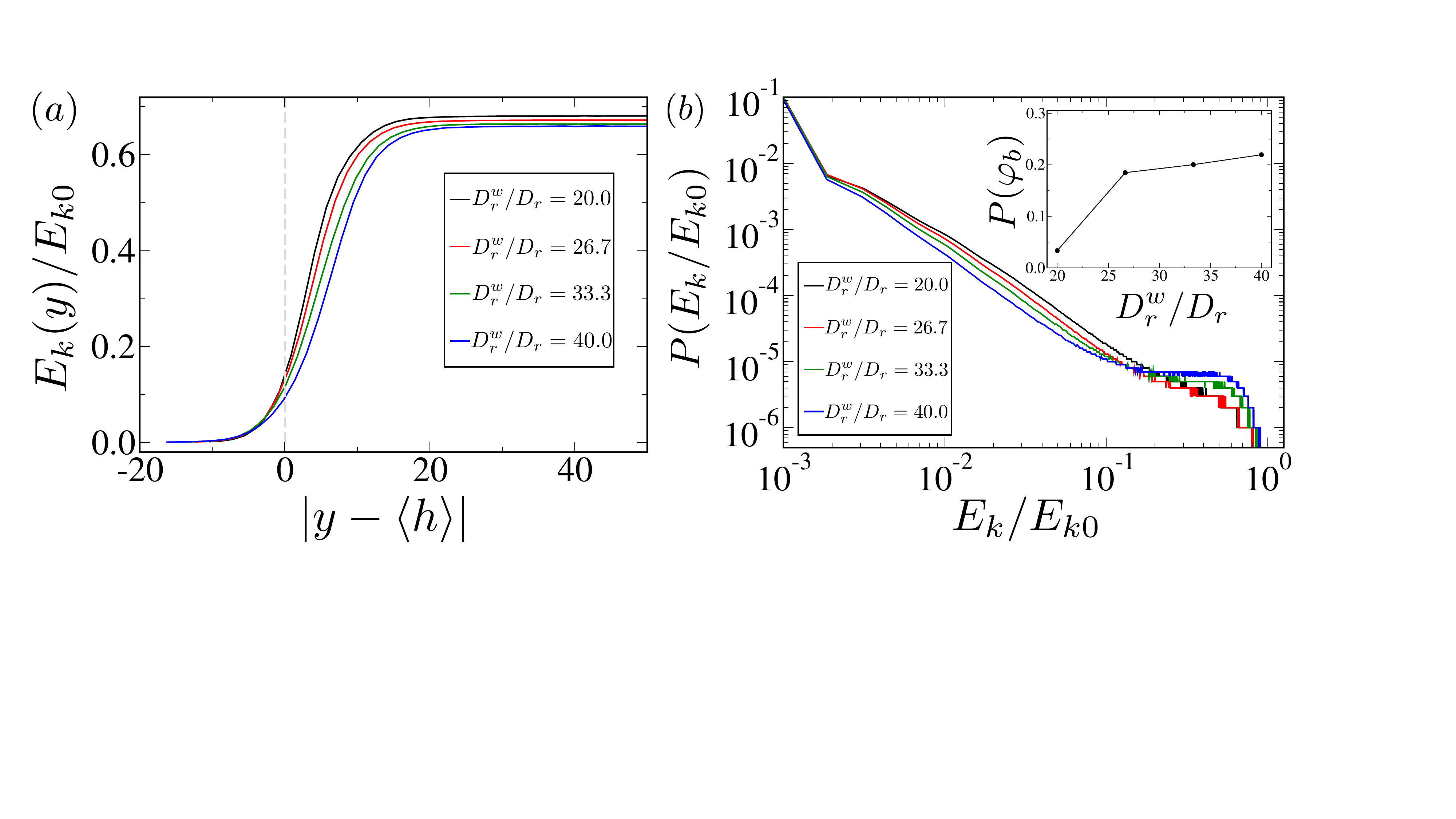}
	\caption{{{(a) Normalized kinetic energy of particles $E_k(y)/E_{k0}$ as a function of distance from the interface $|y- \langle h \rangle|$ for the case of the wet walls, plotted for increasing values of $D_r^w/D_r$ in the symmetric to asymmetric wetting transition regime at Pe =150. The grey line at $y = 0$ indicates the interface location. A slight variation of the energy in the vapor phase is related to the difference in particle densities.
 (b) Distribution of the normalized kinetic energy showing higher weightage for having high kinetic energy for increasing $D_r^w/D_r$. (Inset) Increase in bubble formation with increasing $D_r^w/D_r$ for the case of the wet walls plotted for the symmetric to asymmetric wetting transition regime. This has been shown here as the minimum in the probability distribution of the local density, focusing on the low density region to track bubbles, plotted for $Pe = 150$.}}  
	}
	\label{fig:4}
\end{center}	
\end{figure}

We have observed from the simulations that the dense state often encompasses `bubbles', temporarily formed compact domains with low density 
{(see Fig.~S5).}
Qualitative analyses of simulation results indicate 
an increase in the number of bubbles formed within the dense layer. Such bubble formations are reported inside the dense regions formed by MIPS and in the presence of walls ~\cite{martinroca2021mips, caporusso2020mips, shi2020self, Turci_PRL_2021}, and they can potentially lead to a local enhancement in particle displacement and an overall increase in the effective kinetic energy. We first quantify this effect by calculating the effective kinetic energy as a function of distance from the wall, $\langle E_k(y) \rangle_x = \frac{1}{2}\langle \left(\frac{\Delta {\bf r} (y)}{\Delta t}\right)^{2}\rangle_x$ on which the dense layer is formed. Here, $\langle \rangle_x$ symbolizes averaging over the particles whose vertical distance from the wet wall lies between $y$
 and $y+\delta y$ ($\delta y = 1.6 \sigma$), and we take $\Delta t = 0.1$. This choice of $\Delta t$ ensures that we capture the local dynamics of active particles while they remain within the dense layer. On the other hand, a smaller $\Delta t$ should lead to capturing the effect of translational noise due to velocity fluctuations.
 In Fig.~\ref{fig:4}(a), we plot the scaled effective energy $E_k(y)/E_{k0}$, where $E_{k0} = v_0^2/2$, the kinetic energy corresponding to free self-propulsion, as a function of the distance from the liquid-vapor interface, $y-\langle h \rangle$, where $\langle h \rangle$ is the mean position of the liquid-vapor interface (see {Supplementary Material}). As shown in 
 Fig.~\ref{fig:4}(a), we do not observe any significant effect of {$D_r^w/D_r$} in $\langle E_k \rangle_x$, especially 
 inside the dense layer, $y < \langle h \rangle$. This indicates that the average kinetic energy of the particles in the dense layer is not significantly affected by an increase in {$D_r^w/D_r$}. As a next step, we quantify the probability distribution of the kinetic energy, $p(E_k/E_{k0})$ for the particles within the dense layer, whose vertical distance from the wet wall is confined within the distance to the liquid-vapor interface ({see Supplementary Material}). Interestingly, we find a qualitative difference in this distribution between systems with different {$D_r^w/D_r$}(Fig~\ref{fig:4}(b)). For large {$D_r^w/D_r$},  $p(E_k/E_{k0})$ shows a small peak at large $E_k/E_{k0}$ values, indicating an enhanced particle displacement with \Drw. 
Since the qualitative observation indicates that the frequency of bubble formation also increases with \Drw, it is plausible that the increase in the fraction of particles with high mobility inside the dense region is related to the formation of bubbles. 

To systematically investigate the bubble formation, we quantify the density of bubbles in the dense cluster near the wet walls. For this, we compute the local density distribution $P(\varphi)$ vs. $\varphi$ for the dense cluster up to a distance $y = \langle h \rangle - w$ (see {Supplementary Material}) and examine how $P(\varphi)$ evolves with increasing $D_r^w$ 
{(see Fig.~S6)}
The lower tail of $P(\varphi)$ for $0<\phi<0.005$ should naturally include the regions where bubbles are formed since the peak value of $P(\varphi)$ corresponds to $\varphi \simeq 1.0$. Thus, we define the bubble probability, 
$P(\varphi_b) = \int_0^{\varphi_c} P(\varphi) d\varphi$, where $\varphi_c$ is a threshold density chosen here as $0.005$ as the lowest non-zero density tracked. We observe a systematic increase in $P(\varphi_b)$ with an increase in {$D_r^w/D_r$} (see Fig.~\ref{fig:4}(b)-inset), indicating enhanced bubble formation. Thus, the simultaneous existence of enhanced bubble formation and an increase in the fraction of particles with large $E_k$ hints at a causal connection between these two phenomena. Although an increase in bubble formation does not significantly affect the mean kinetic energy of the particles in the dense state, we observe a qualitative difference in the energy distribution. These high-energy particles add energy fluctuations that can possibly trigger the destabilization of the entire dense state. 
\subsection{Characterizing the wetting transitions}
\begin{figure}
  \begin{center}
     	\includegraphics[width=0.85\columnwidth]{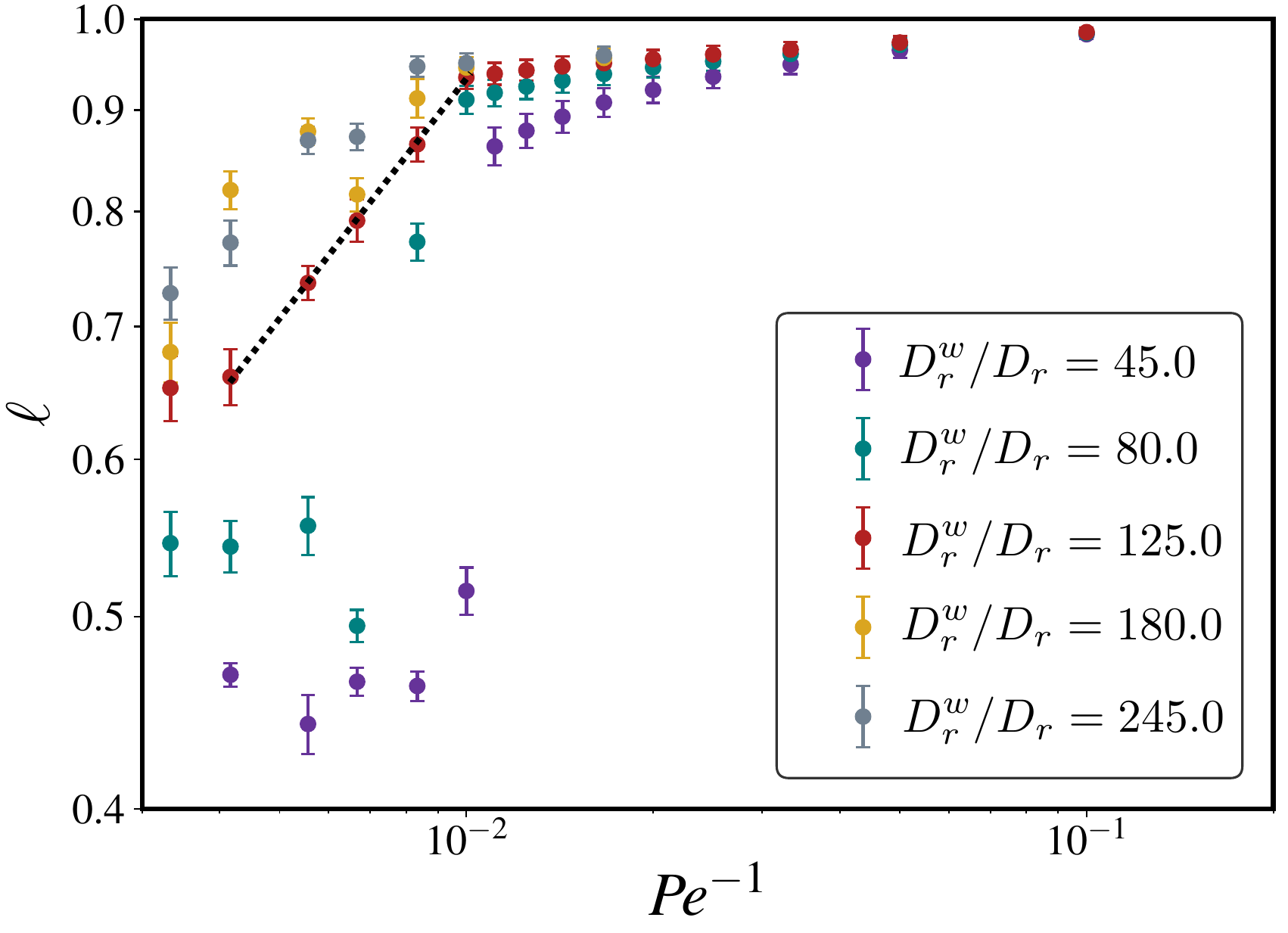}
 	\caption{{Variation of the average vacancy fraction $\langle \ell \rangle$ as a function of the inverse activity. The dotted line represents a linear fit of the form $f(x) = ax+b$, with $a = 46$ and $b = 0.48$. }
 	}
 	\label{fig:5}
 \end{center}	
 \end{figure}

Using the simulation data, we now characterize the type of wetting transition, which is crucial for our understanding of this phenomenon. For equilibrium wetting transitions on a solid-liquid interface, both first-order and continuous transitions have been predicted and observed, depending on the types of interactions between the components and the system temperature~\cite{cahn1977critical,nakanishi1982multicriticality,wang2001line, murata2010surface}. For active systems, an appropriate order parameter has to be defined to track such transition properties. 
 Sepúlveda and Soto have studied the wall aggregation of active particles on inert walls as a function of particle reorientation rate in the bulk~\cite{sepulveda2017wetting, SepulvedaSoto_PRE2018_activewetting}, and the instantaneous fraction of empty sites on the wall has been used as the order parameter. 
 Similarly, we use the vacancy fraction  $\langle \ell \rangle$  as the order parameter here, which has been calculated by averaging the vacancy fraction over time and for multiple realizations for a given set of parameters. As was done in previous studies, we compute $\langle \ell \rangle$ as a function of $Pe$, which is a measure of active particle persistence. In Fig.~\ref{fig:5}, we plot $\langle \ell \rangle$ vs $Pe^{-1}$ for the range of  {$D_r^w/D_r$} where partial wetted states are also observed. 
 For a  small Pe (large Pe$^{-1}$) the system displays a complete dry state as $\langle \ell \rangle \rightarrow 1$ and for larger Pe (smaller Pe$^{-1}$), $\langle \ell \rangle$ is between 0 and 1. The transition is rather discontinuous for relatively small {$D_r^w/D_r$}, as the change in $\langle \ell \rangle$ is rather abrupt. However, when $D_r^w$ is increased, we observe signs of a continuous transition from $\langle \ell \rangle <1$ to $\langle \ell \rangle \simeq 1$. In this range, $\langle \ell \rangle$ even indicates a power-law-like behavior as a function of $Pe$. For {$D_r^w/D_r \simeq 125$}, $\ell$ vs $Pe^{-1}$ approximately fits to a power-law, $\ell \sim Pe^{-\alpha}$, where $\alpha \simeq 1$ (see Fig.~\ref{fig:5}). Interestingly, the exponent is consistent with previous observations on wetting transitions in active systems~\cite{sepulveda2017wetting,SepulvedaSoto_PRE2018_activewetting}. However, the transition that we observe is from a partial wetted to a complete dry state (drying transition), unlike shown in reference~\cite{sepulveda2017wetting}.
 In those studies, instead of changing the wall properties, the wetting transitions were observed by changing the particle reorientation rate. Therefore, this observation is indicating the existence of a universal behavior in wetting transitions achieved through various mechanisms. 

\section{Summary and discussions}

In this study, we numerically demonstrate distinct morphological transitions of active aggregates as the wall-particle interactions are modified. The modification of such wall interactions is not brought in through any change in the interaction potentials but through an enhancement in the reorientation rate of the particles in the vicinity of the wall. This modification resembles the observations in bacterial systems in which the individual bacteria actively modify its reorientation rate near solid boundaries. Therefore, this model depicts such active modification of wall-particle interactions. The morphological states observed in this non-equilibrium system show similarities to wetting transitions in equilibrium liquid-solid interfaces. We obtain droplet formation of active aggregates, resembling partial wetting in equilibrium fluids. The parameter space spanned by the activity and noise enhancement at the wall reveals the region in which partial wetting states are observed. We also study the dependence of contact angle in the partial wetting regime on both particle motility and wall reorientation rates, along with the effect of wall reorientation rate on the overall structure and dynamics of the dense state. We find that enhanced reorientation increases the number of particles with higher kinetic energy. Also, we find an increase in the occurrence of bubbles inside the dense clusters. Since the particle mobility can be higher in the vicinity of bubbles, these two observations are likely to be causally connected.   

While the wetting transitions have previously been studied in the context of active systems~\cite{sepulveda2017wetting, Turci_PRL_2021}, these results add another axis to the relevant parameter space, namely the active wall interaction. We study the type of transitions by measuring the order parameter, namely the wall vacancy fraction. We observe an intermediate $D_r^w$ where the transition shows signatures of a continuous one as a function of an increase in the reorientation rate. Interestingly the exponent of this continuous transition agrees to previously observed values~\cite{sepulveda2017wetting}.
Our previous studies have shown similar morphological transitions in the presence of porous walls~\cite{Das2020_SM, Das2020_PRE}. However, the present study reveals that active morphological transitions have a broader scope, with porosity not being the sole driving factor. There, the cross-wall particle flux induced by wall porosity possibly adds an enhanced noise at the wall, leading to morphological transitions in such systems.


The study by Rogers et al. \cite{Rogers_PRL_2012} on the rotational diffusion of colloids near surfaces offers valuable experimental inspiration. Recently developed self-propelled robots can be used to validate our numerical findings~\cite{paramanick2024programming}. Another potential experimental application of our findings could involve bacteria confined within rigid walls treated with chemical repellents. Adjusting the repellent strength would influence bacterial movement away from the wall, providing a controlled way to study wall-particle interactions. 

{Our study on the wetting behavior of active Brownian particles can be extended in many different directions. For example, anisotropic or rod-like self-propelled objects are expected to have a different wetting-like behavior which can be systematically explored~\cite{wensink2008aggregation,elgeti2009self,khan2022effect}. Further, the wall aggregation of deformable particles~\cite{loewe2020solid,ohta2009deformable}, and active ring polymers~\cite{mousavi2019active, theeyancheri2024dynamic} can also be systematically explored. 
This study can be repeated for different types of walls with varying curvature, roughness, geometry, and in the presence of external fields~\cite{Bechinger2016, pincce2016disorder, shaik2023confined}. Our study is conducted for a fixed area fraction (or a fixed number of particles), and the morphological changes are induced due to the change in wall interaction. One pertinent question is how this wetting behavior will be influenced if the number of particles is not constant, in a grand canonical arrangement~\cite{neta2021wetting}.
}



\section*{Supplementary Material}
{For list of parameters, supplementary figures and associated text, movie captions, refer to supplementary material.}


{\it Acknowledgement}
We acknowledge SpaceTime2 HPC facility at IIT Bombay and computing support from the Max Planck Computing and Data Facility. RC acknowledges financial support from SERB India via research grant no. CRG/2021/002734. RC and SD acknowledge helpful discussions with Dibyendu Das, Amitabha Nandi, Anirban Sain and Mithun Mitra. SD acknowledges helpful discussions with Ricard Alert, Frank Jülicher, and Debasish Chaudhuri.  

\noindent{\bf Author declarations}\\
{\it Conflict of interest:} The authors have no conflicts to disclose.\\

\noindent{\it Author contributions}- SD and RC: Conceptualization (equal); developing the numerical model (equal); running the simulations (equal); Analyzing the data (equal); Writing the manuscript (equal).\\

\noindent{\bf Data availability}\\
The data that support the findings of this study are available from the corresponding author upon reasonable request.

\newpage
\begin{appendix}

\title{Supplemental Material for ``Active wetting transitions induced by rotational noise at solid interfaces''}

\maketitle
\renewcommand{\thefigure}{S\arabic{figure}}
\setcounter{figure}{0}  

\section{List of symbols}\label{app:list}
\begin{table}[h]
\centering
\begin{tabular}{|p{0.1\textwidth}|p{0.35\textwidth}|} 
    \hline
    \textbf{Parameter} & \textbf{Definition} \\ [0.45ex] 
    \hline
    ${\bf r}_i$ & Particle position\\
    \hline
    $\hat{\bf e}_i$ & Particle polarity\\
    \hline
    $\phi_i$ & Orientation angle of the polarity vector\\
    \hline
    $\sigma$ & WCA interaction range\\
    \hline
    $D$ & Diffusion coefficient\\
    \hline
    $D_r$ & Rotational diffusion coefficient in bulk\\
    \hline
    $D_r^w$ & Rotational diffusion coefficient near walls\\
    \hline
    $Pe$ & P\'eclet number \\ 
    \hline
    $N$ & Number of particles\\
    \hline
    $\varphi$ & Area fraction\\
    \hline
    $\rho$ & Local density\\
    \hline
    $v_0$ & Particle motility\\
    \hline
    $\ell$ & Vacancy fraction averaged\\ & over all realisations\\
    \hline
    $\ell_w$ & Vacancy fraction averaged\\& over wet walls, except\\& for parameters which exhibit\\& complete drying for all realisations\\
    \hline
    $\theta$ & Contact angle of liquid drop\\
    \hline
\end{tabular}
\caption{Parameter definitions.}
\label{Tab:1}
\end{table}

\section{Steady-state density distribution and Vacancy fraction}\label{app:density}

To characterize the steady-state density distribution, we divide the system into $b$ square bins and measure the time-averaged number of particles in each bin, $\langle n_b \rangle$. The two dimensional density distribution is then given by $\rho(x,y) = \langle n_b \rangle\pi \sigma^2/(4 A_b)$, where $A_b$ denotes the area of the bin, chosen as $A_b = 2.0 r_c$. We plot this in 
Fig.~1(b)
for a typical steady-state configuration. For small {$D_r^w/D_r$}, we observe high local density near both the walls as shown in 
Fig.~1(b)(i)
corresponding to a {{symmetric complete wetting}} state. Increasing {$D_r^w/D_r$}, we observe high local density only on one wall, as shown in 
Fig.~1(b)(ii).
This transition occurs around {$D_r^w/D_r = 45$}, corresponding to an {{asymmetric wetting state}}. Further increase of {$D_r^w/D_r$}, we observe droplet formation as shown in 
Fig.~1(b)(iii)
for {$D_r^w/D_r$ = 125}, corresponding to a {{partial wetting}} state. 

\begin{figure}[h]
    \centering
	\includegraphics[width=0.65\columnwidth]{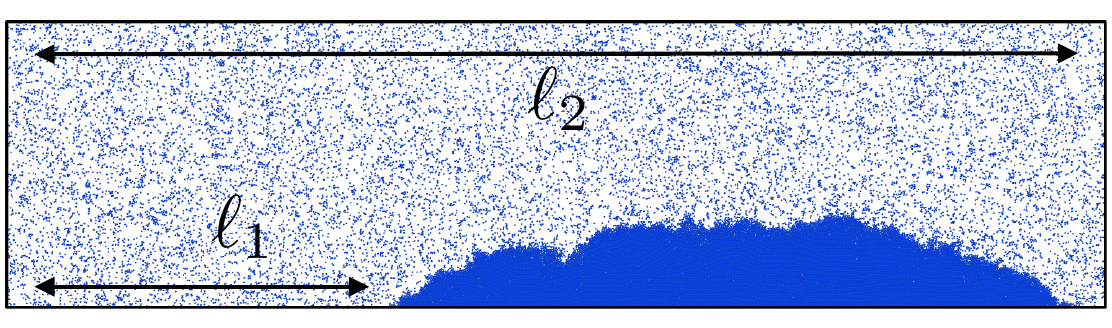}
	\caption{Schematic for calculating vacancy fraction $\ell = (\ell_1 + \ell_2)/2$.
	}
	\label{fig6}
\end{figure}

\begin{figure}[h]
    \centering
	\includegraphics[width=0.98\columnwidth]{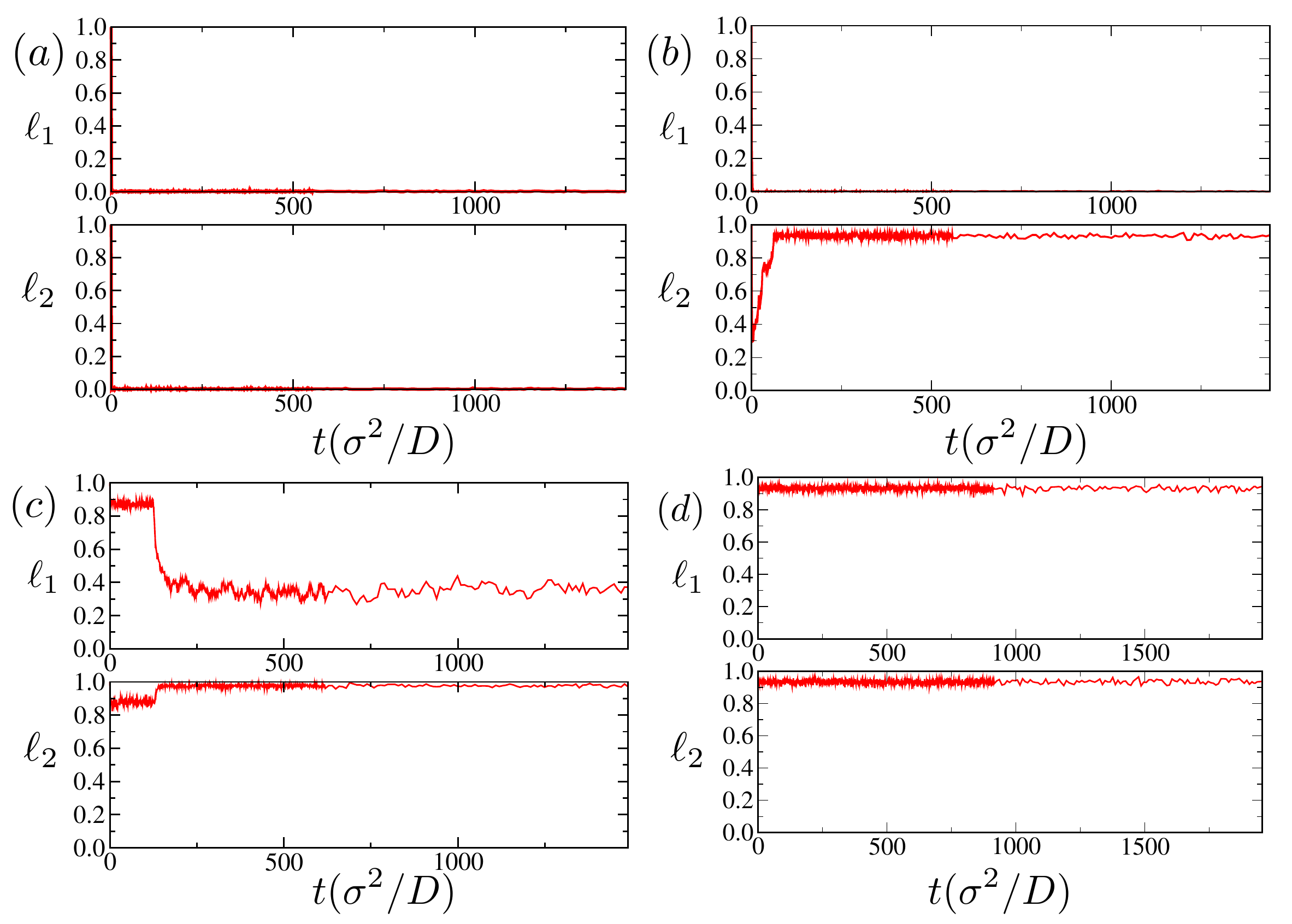}
	\caption{Example of vacancy fraction for (a) symmetric complete wetting state {($D_r^w/D_r = 10$)}, (b) asymmetric complete wetting {($D_r^w/D_r = 45$)}, (c) partial wetting {($D_r^w/D_r = 125$)} and (d) complete drying state {($D_r^w/D_r = 320$)} with $N= 38400$, $Pe = 240$.
	}
	\label{fig7}
\end{figure}

To quantify the different morphological states, we calculate the vacancy fraction $\ell$. We bin the system along the $x$-axis and calculate the number of unoccupied bins, normalized by the system length along the $x$ axis. From this, we calculate the fraction of vacant length along the wall and take the mean over both walls. This quantity is defined as the vacancy fraction, $\ell = (\ell_1 + \ell_2)/2$, where $\ell_1$ and $\ell_2$ are the vacancy fraction for the lower and upper wall, respectively, as shown in Fig.~\ref{fig6}. In Fig.~\ref{fig7}, we show typical examples for the four different wetting states.

\section{Cluster fraction}\label{app:clusterfraction}

 We consider two particles to be part of a cluster if they are within the interaction cutoff range, $2^{1/6}\sigma$. For all the parameter values that were analyzed, the aggregation takes place near the wall. We calculate the total number of particles that are part of the dense phase clustered around the wall $N_c$ and define the cluster fraction as $N_c/N$, where $N$ is the total number of active particles in the system. We calculate this as a time-averaged quantity over steady states averaged over independent runs.

\section{Bi-stability statistics}\label{app:bistability}

\begin{figure}[h]
    \centering
	\includegraphics[width=0.98\columnwidth]{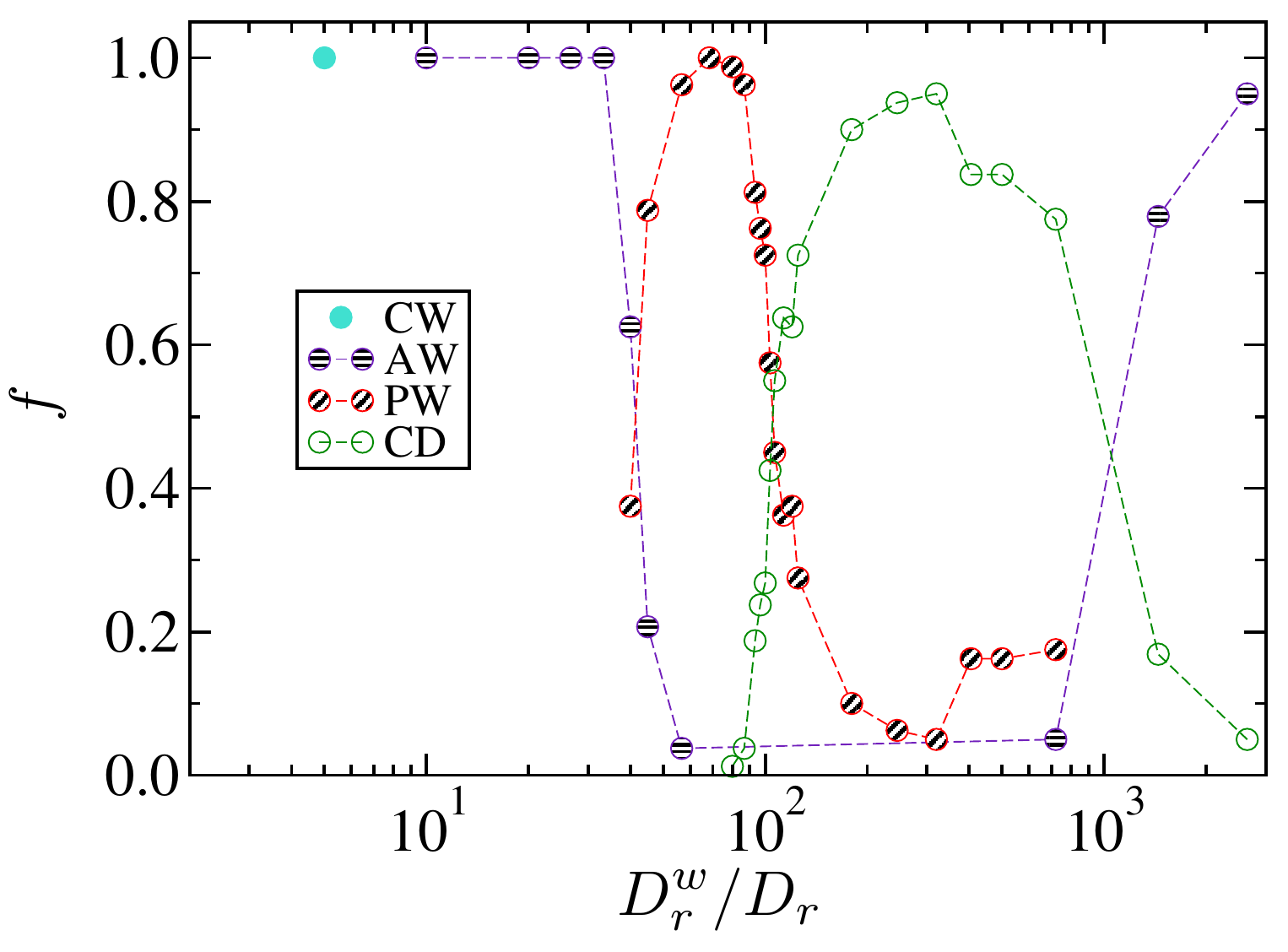}
	\caption{Fraction of runs showing the different wetting states, with transition regions showing bistability, characterized here for a system having $N = 8008$ particles for $Pe = 150$, for $80$ independent runs.
	}
	\label{fig:8}
\end{figure}

We quantify the number of independent runs showing the different wetting states, where the transition regions between the states show indications of bistability, as shown in Fig.~\ref{fig:8}. 
We note that for some parameters, we have observed late-time transitions from one morphological state to another. Thus, the characterization of the morphological states is based on the transitions observed until the end of the simulation. 

\section{Estimation of contact angle and its variation  with activity}\label{app:contactangle}
We quantify the contact angle from the average density distribution $\rho(x,y)$ for the dense clusters formed at the walls. First we identify the liquid-gas interface where the density becomes negligible, $\rho(x,y) \simeq 0$. The angle formed by intersecting the local tangent of the interface at $y=0$, with the wall provides contact angle, $\theta$. In the steady state, the mean contact angle is calculated by averaging $\theta$ over the steady-state configurations.

\begin{figure}[h]
	\includegraphics[width=0.98\columnwidth]{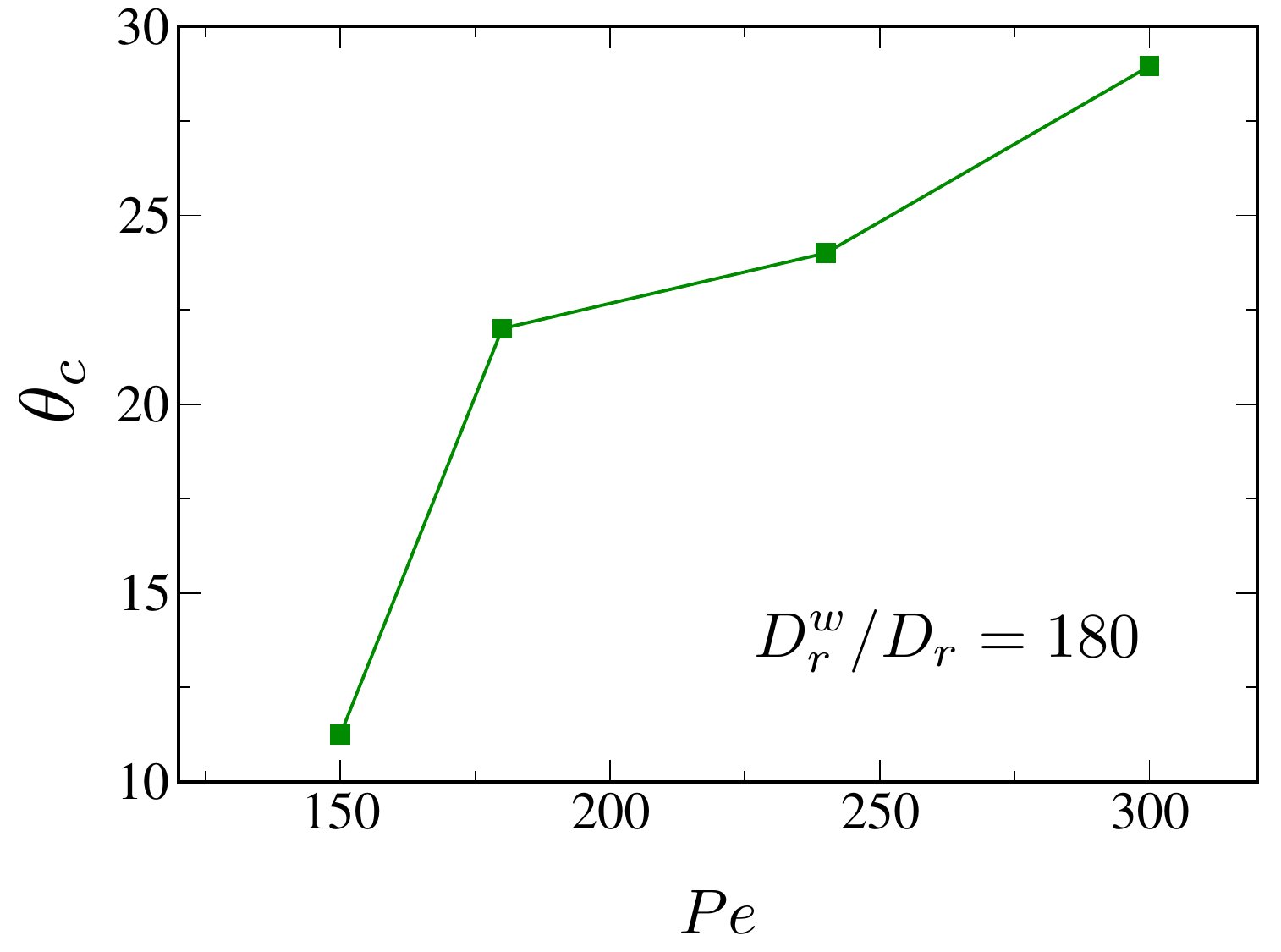}
	\caption{Variation in contact angle of droplet with activity for {$D_r^w/D_r = 180$}, corresponding to 
    Fig. 3(a).
	}
	\label{fig:9}
\end{figure}

To quantify the effect of activity on droplet morphology, we calculate the contact angle $\theta$ as a function of activity $Pe$. Our analysis confirms that $\theta$ increases with activity, as shown in Fig.~\ref{fig:9}, consistent with the increase in interfacial curvature.

\section{Kinetic energy distribution}\label{app:kedist}
 We calculate the kinetic energy distribution $P(E_k)$, where we normalize the count of particles in each bin by the total number of particles $N$ and the time $t$. We quantify this by measuring the kinetic energy $E_k = \frac{1}{2}(\frac{\Delta {\bf r}}{\Delta t})^{2}$ close to the walls ($|y - y_w| < 10.0\sigma$). 

\section{Quantification of bubbles}\label{app:bubbles}
To quantify the bubbles as shown in Fig.~\ref{fig:10} in the wet wall, we locate the dense cluster formed along the wall, using the method described in Appendix~\ref{app:clusterfraction}. Using the data, we define the local heights $h(x,t)$ for the dense cluster, which is the maximum $y$ distance of the particle at the dense cluster edge from the wall. We can then measure the mean height as $\langle h \rangle = {1\over{(T-t_1)}}{1\over L_x}\int_{t_1}^{T}\int_0^{L_x}h(x,t) dx\:dt$, in the steady state.
Further, we calculate the steady-state interfacial fluctuation as $ w^2 = {1\over{(T-t_1)}}{1\over L_x}\int_{t_1}^{T}\int_0^{L_x}(h(x,t) - \langle h \rangle)^{2} dx\:dt$
We compute the local density for region up to the distance $y = \langle h \rangle - w$ from the wet wall. To estimate the bubble formation, we focus on the lowest local density.  

\begin{figure}[h]
	\includegraphics[width=0.98\columnwidth]{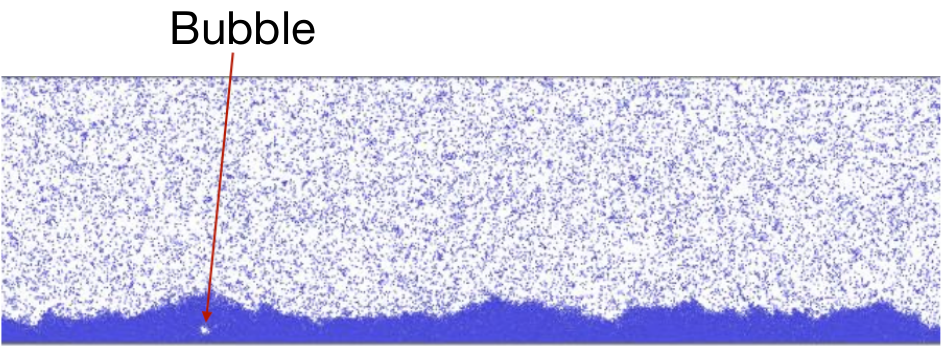}
	\caption{Typical snapshot of observed bubble formation, shown here for {$D_r^w/D_r = 40$}, $Pe = 150$.
	}
	\label{fig:10}
\end{figure}

\begin{figure}[h]
	\includegraphics[width=0.98\columnwidth]{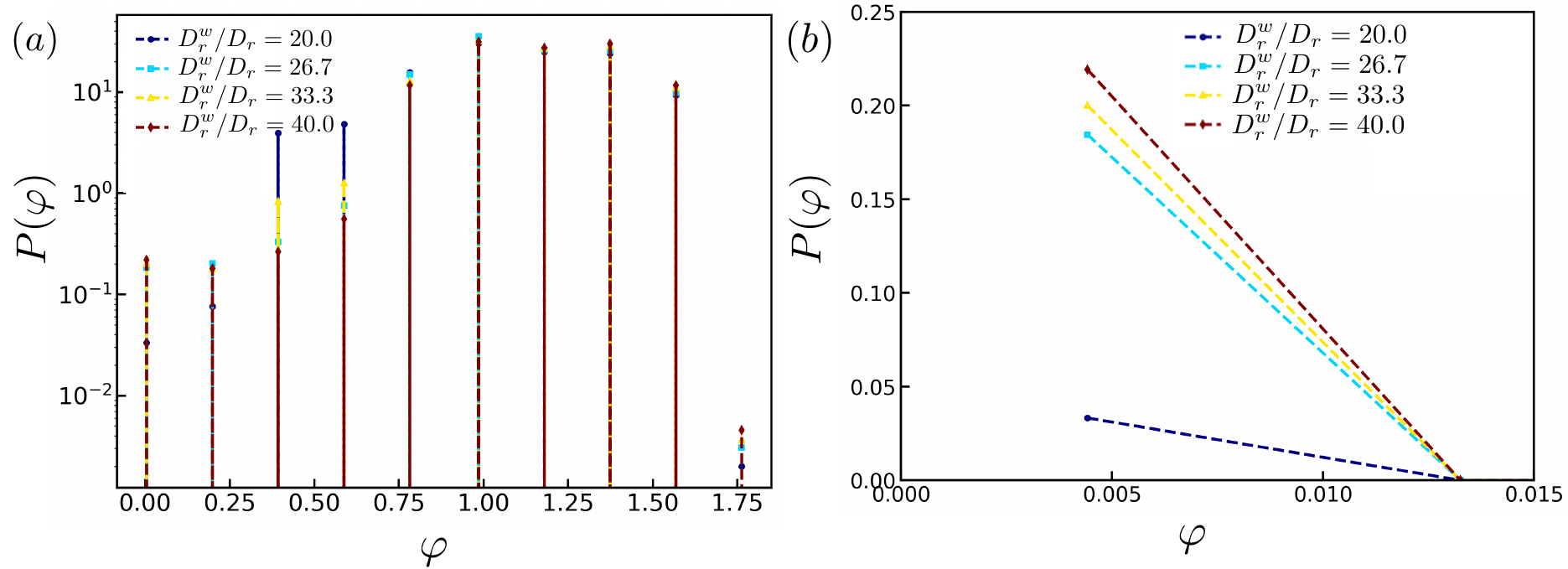}
	\caption{(a) Local density distribution $P(\varphi)$ vs. $\varphi$ for increasing {$D_r^w/D_r$} plotted in log-linear scale. $P(\varphi)$ follows the normalization $\sum_i p(\varphi_i)\Delta \varphi_i =1 $. (b) Zoomed in view highlighting the lower tail of $P(\varphi)$, plotted on a linear-linear scale. 
	}
	\label{fig:11}
\end{figure}

\section{Noninteracting particles}

\begin{figure}
    \centering
    \includegraphics[width=0.98\linewidth]{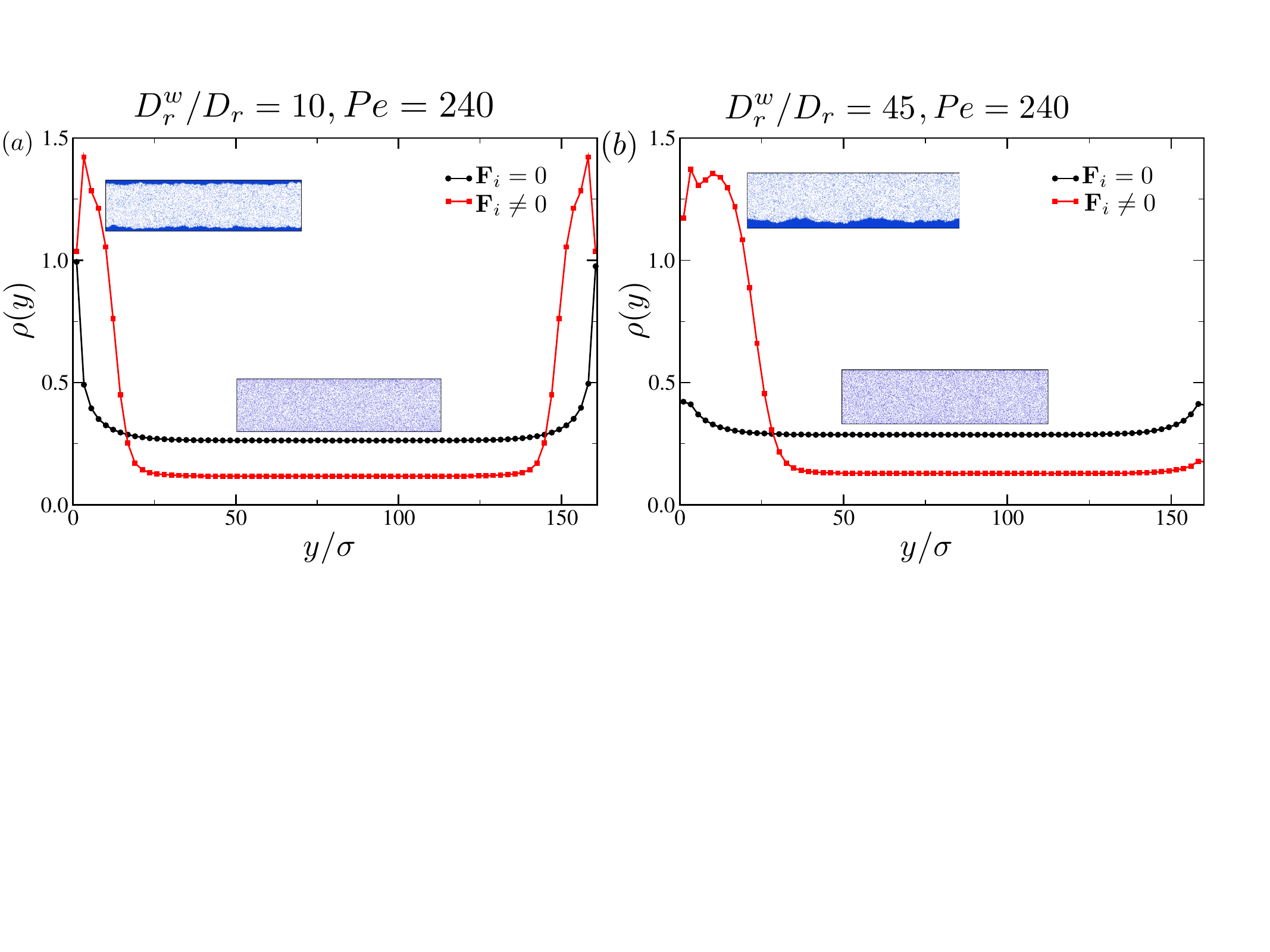}
    \caption{Comparison of density profiles with and without particle interaction, for (a) symmetric wetting case $D_r^w/D_r = 10$, $Pe = 240$, (b) asymmetric wetting case $D_r^w/D_r = 45$, $Pe = 240$.}
    \label{fig:12}
\end{figure}

{We have conducted simulation of active Brownian particles in the absence of interparticle interactions and compared with the results obtained for interacting ABPs. Fig.~\ref{fig:12} shows the density profiles in the absence of particle interaction ${\bf{F}}_i = 0$, for the parameters in which we observe (a) symmetric, and (b) asymmetric complete wetting for interacting particles. The particle aggregation still occurs even in the absence of interparticle interactions. However, we do not observe an asymmetric complete wetting state in this case, as the liquid-vapour interface is absent.}

\section{Comparison of the area fraction}

\begin{figure}
    \centering
    \includegraphics[width=0.9\linewidth]{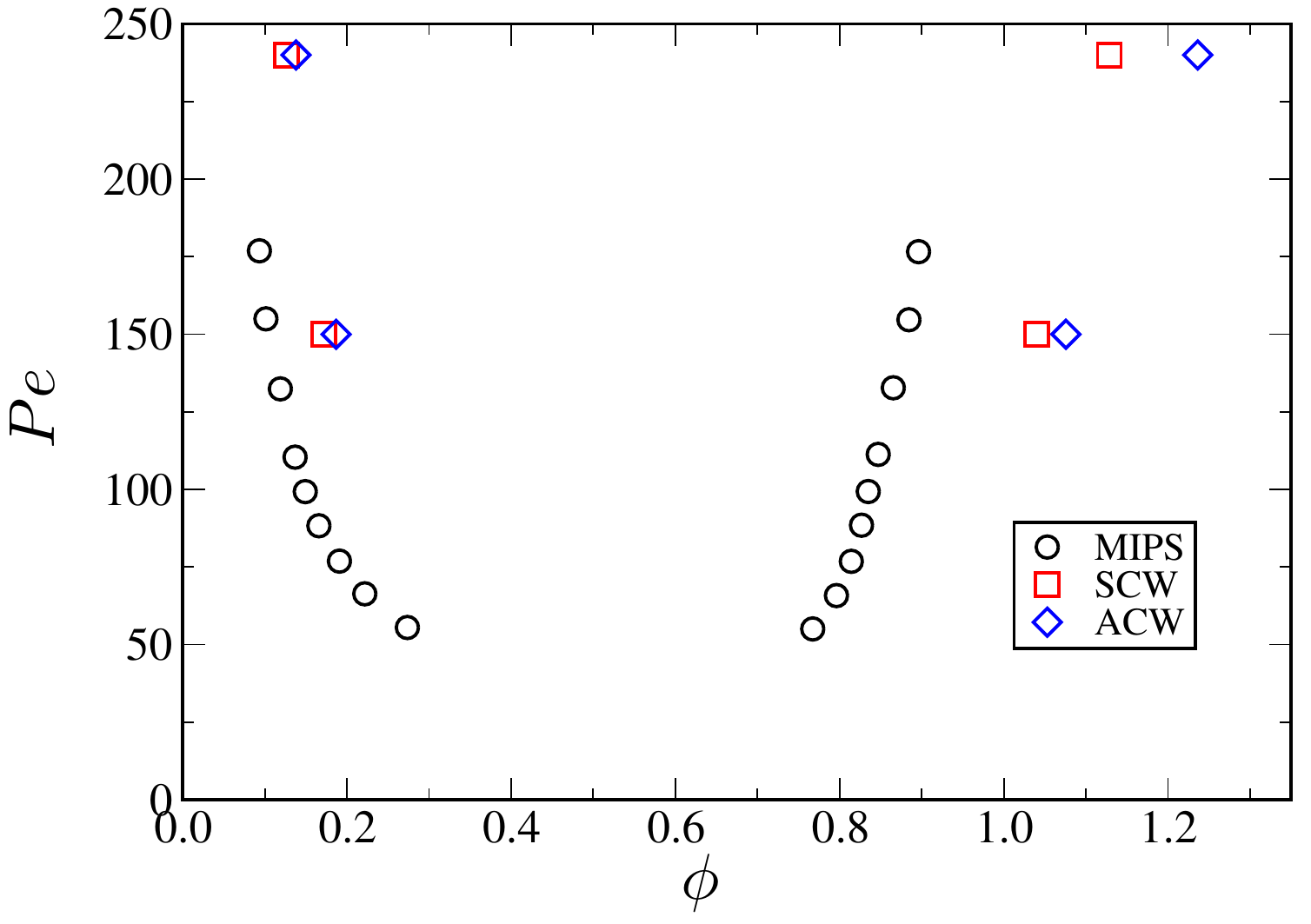}
    \caption{{Comparison of area fractions of the liquid and vapor states in (i) the case of active Brownian particles in periodic boundary conditions (data extracted from Siebert et al.~\cite{Siebert2018}., (ii) SCW state ($D_r^w/D_r = 10$, $Pe = 150$ and $Pe = 240$), and (iii) ACW state ($D_r^w/D_r = 45$, $Pe = 150$ and $Pe = 240$)}}
    \label{fig:13}
\end{figure}

{We compare the area fractions observed in the dense and vapor states in both the symmetric and asymmetric complete wetting states, and contrast these with the case of active Brownian particles (ABPs) undergoing motility-induced phase separation (MIPS), as shown in Fig.~\ref{fig:13}. To distinguish between the liquid and vapor phases, we analyze the mean interface height, $\langle h \rangle$, which quantifies the location of the liquid-vapor interface. We then compute the averaged density profiles separately for the dense and vapor phases and compare them with the corresponding profiles for ABPs undergoing MIPS, as reported in reference \cite{Siebert2018}.}

\section{Supplementary movies}
\begin{itemize}
    \item MOVIE1: Typical cluster morphology observed for symmetric complete wetting (SCW) state, observed at $Pe = 240$ and {$D_r^w/D_r$ = 10}. 
    \item MOVIE2: Typical cluster morphology observed for asymmetric complete wetting (ACW) state, observed at $Pe = 240$ and {$D_r^w/D_r$ = 45}.
    \item MOVIE3: Typical cluster morphology observed for partial wetting (PW) state, observed at $Pe = 240$ and {$D_r^w/D_r$ = 125}.
\end{itemize}


\end{appendix}
\bibliographystyle{unsrt}
\bibliography{reference}

\end{document}